\documentclass[aps,twocolumn,preprintnumbers,amsmath,amssymb,prb,superscriptaddress]{revtex4}
\usepackage[dvips]{epsfig}
\usepackage{subfigure}
\usepackage{graphicx}
\usepackage{longtable}
\usepackage[T1]{fontenc}
\usepackage[ansinew]{inputenc}
\usepackage{ulem}
\usepackage{color}

\begin{document}

\newsavebox{\smlmat}% Box to store smallmatrix content
\savebox{\smlmat}{$\left(\begin{smallmatrix}7&1\\2&5\\\end{smallmatrix}\right)$}

\title{Vertical bonding distances and interfacial band structure of PTCDA on a Sn-Ag surface alloy}

\author{Johannes Seidel}
\email[]{jseidel@rhrk.uni-kl.de}
\affiliation{Department of Physics and Research Center OPTIMAS, University of Kaiserslautern, Erwin-Schroedinger-Strasse 46, 67663 Kaiserslautern, Germany}
\author{Leah L. Kelly}
\affiliation{Department of Physics and Research Center OPTIMAS, University of Kaiserslautern, Erwin-Schroedinger-Strasse 46, 67663 Kaiserslautern, Germany}
\author{Markus Franke}
\affiliation{Peter Gr\"unberg Institut (PGI-3), Forschungszentrum J\"ulich, 52425 J\"ulich, Germany}
\affiliation{J\"ulich-Aachen Research Alliance (JARA) -- Fundamentals of Future Information Technology, 52425 J\"ulich, Germany}
\author{Christian Kumpf}
\affiliation{Peter Gr\"unberg Institut (PGI-3), Forschungszentrum J\"ulich, 52425 J\"ulich, Germany}
\affiliation{J\"ulich-Aachen Research Alliance (JARA) -- Fundamentals of Future Information Technology, 52425 J\"ulich, Germany}
\author{Mirko Cinchetti}
\affiliation{Experimentelle Physik VI, Technische Universit\"at Dortmund, 44221 Dortmund, Germany}
\author{Martin Aeschlimann}
\affiliation{Department of Physics and Research Center OPTIMAS, University of Kaiserslautern, Erwin-Schroedinger-Strasse 46, 67663 Kaiserslautern, Germany}
\author{Benjamin Stadtm\"uller}
\affiliation{Department of Physics and Research Center OPTIMAS, University of Kaiserslautern, Erwin-Schroedinger-Strasse 46, 67663 Kaiserslautern, Germany}
\affiliation{Graduate School of Excellence Materials Science in Mainz, Erwin Schroedinger Strasse 46, 67663 Kaiserslautern, Germany}

\date{\today}

\begin{abstract}
Molecular materials enable a vast variety of functionalities for novel electronic and spintronic devices. The unique possibility to alter or substitute organic molecules or metallic substrates offers the opportunity to modify and optimize interfacial properties for almost any desired field of application. For this reason, we extend the successful approach to control molecular interfaces by surface alloying. We present a comprehensive characterization of the structural and electronic properties of the interface formed between the prototypical molecule PTCDA and a Sn-Ag surface alloy grown on an Ag(111) single crystal surface. We monitor the changes of adsorption height of the surface alloy atoms and electronic valence band structure upon adsorption of one layer of PTCDA using the normal incidence x-ray standing wave technique in combination with momentum-resolved photoelectron spectroscopy. We find that the vertical buckling and the surface band structure of the SnAg$_2$ surface alloy is not altered by the adsorption of one layer of PTCDA, in contrast to our recent study of PTCDA on a PbAg$_2$ surface alloy [Phys. Rev. Lett. 117, 096805 (2016)] . In addition, the vertical adsorption geometry of PTCDA and the interfacial energy level alignment indicate the absence of any chemical interaction between the molecule and the surface alloy. We attribute the different interactions at these PTCDA/surface alloy interfaces to the presence or absence of local $\sigma$-bonds between the PTCDA oxygen atoms and the surface atoms. Combining our findings with results from literature, we are able to propose an empiric rule for engineering the surface band structure of alloys by adsorption of organic molecules. 
\end{abstract}
\pacs{...}
\maketitle
\section{Introduction}
In the fast growing field of organic electronics and spintronics, it was soon realized that the device relevant properties of molecular assemblies on surfaces are determined by the intrinsic properties of the active molecular materials and the metallic substrates as well as by the phenomena occurring at the interfaces between both materials. While the most important intrinsic properties of molecular materials and metallic substrates are rather well understood today \cite{Gutmann1967,SPANGGAARD2004125,Ishiiadvancedmaterials1999,Schwarze1446,C3CS60449G,TAUTZ2007479}(and references therein), the interfacial properties of metal-organic (hybrid) interfaces are still subject of intense investigations. This is mainly due to the large complexity of such interfaces, which is rooted in the delicate interplay between different types of molecule-surface and intermolecular interactions that can occur in such systems \cite{TAUTZ2007479,SEKI2001298,Hofmann_2013}. These interactions can result in a severe modification of both sides of the metal-organic interfaces and, in selected cases, in the emergence of exotic phenomena such as adsorption-induced magnetic order in otherwise diamagnetic surfaces\cite{MaMari2015}. 

For the metal side, the molecule-surface interaction can lead to a depopulation (shift) or a suppression of the metallic surface states upon the adsorption of organic molecules \cite{ZHU20041,SchwalbPRL2008, Marksprb2011}. On the molecular side, weak chemisorption of organic complexes on metallic surfaces can result in a charge redistribution across the interface typically coinciding with charge transfer from the surface bands into the former lowest unoccupied molecular orbital (LUMO) \cite{DUHM2008111,Kroger.2010,Zou.2006} . Thereby, the amount of charge transfer into the molecular layer crucially depends in the interfacial interaction strength \cite{Koch_2008,ishii1997}. For even stronger chemical interaction between molecules and surfaces, new states can be formed due to the hybridization between molecular and metallic states. These so called hybrid interface states are most frequently observed on highly reactive transition metal surfaces \cite{hybridinterface_lach,hybridinterface_shi,Cinchettinatmat2017}.

In the past, a variety of different approaches has been demonstrated to control the interfacial properties of metal-organic interfaces, for instance by the formation of heteromolecular structures\cite{Stadtmuller.2014,PRBHaming}, metal-organic networks\cite{Evans_2002,Goiri.2016}  or by alkali metal doping\cite{PRL_Cinchetti2010,DING20101786,Akaike.2019}.  In this work, we follow an alternative route and continue our recent approach of controlling the interaction strength across metal-organic interfaces by surface alloying. On noble metal surfaces, surface alloys can be fabricated with extremely high quality by replacing every third surface atom with a heavy metal alloy atom such as Bi, Pb, Sb or Sn\cite{AstPRL2007}. This class of low dimensional binary materials allows one to design the geometric and electronic  properties as well as the surface reactivity of noble metal surfaces by the right choice of the alloy atom. For instance, the vertical buckling of the surface atoms can be varied from $0.10\,$\AA\ to $0.65\,$\AA\ by selecting the appropriate alloy atom \cite{GierzPRB2010}. In a similar way, the electronic band structure of surface alloys can be tuned by material design. All surface alloys of heavy metal atoms on noble metal surfaces exhibit a hole-like hybrid state with a Rashba-type spin splitting. Thereby the dispersion, energy position and spin-splitting of the hole-like hybrid state can be tailored by the chemical properties of the alloy atoms as well as the strength of their atomic spin-orbit coupling \cite{Bentmann_2009,pacile2006,AstPRL2007,Moreschini_2009}. Only recently, even ferromagnetic order has been observed in surface alloys formed between noble metal host materials and rare earth alloying atoms by Ormaza et al.\cite{Ormaza.2016}, making this class of 2d materials even more interesting for application.
%Following this approach, Ormaza et al. recently discovered ferromagnetic order in surface alloys formed between noble metal hosts and rare earth atoms\cite{Ormaza.2016}.  

Here, we will take advantage of the high tuneablity of heavy-metal, noble-metal surface alloys and explore the interfacial properties of the prototypical molecule PTCDA on a SnAg$_2$ surface alloy. Using a combination of surface sensitive techniques such as core-level spectroscopy, normal-incidence x-ray standing waves and momentum-resolved photoemission, we observe two structurally inequivalent adsorption sites of PTCDA on the SnAg$_2$/Ag(111) surface alloy, in agreement with recent findings for PTCDA on the PbAg$_2$/Ag(111) surface alloy  \cite{BSPRB2016,BSPRL2016}. Most interestingly, in the present system there is no charge transfer into the former LUMO of PTCDA, whichshows a discrepancy with the findings for PTCDA/Ag(111) \cite{HauschildPRB2010}  as well as PTCDA on the PbAg$_2$ surface alloy\cite{BSPRB2016,BSPRL2016}. Our experimental evidence points to a clear reduction of the molecule-surface interaction strength between the PTCDA and the Ag(111) surface by the implantation of Sn into the surface layer. This clearly underlines the potential of surface alloying to control the interaction strength and thereby the geometric, electronic and spin-dependent properties of metal-organic hybrid interfaces.

\section{Experimental details}
\subsection{Sample preparation}
All experiments have been carried out under ultra-high vacuum (UHV) conditions at a base pressure of 5$\times$10$^{-10}\,$mbar at room temperature. The Ag(111) single crystals have been cleaned by multiple cycles of Ar-Ion sputtering at various ion energies of 0.5 - 2.0 keV, I$_{drain}=8\,\mu$A and subsequent sample annealing for 20 minutes up to $730\,$K. The (chemical) cleanness of the sample surface was confirmed by the existence and linewidth of the Shockley surface state and by core level spectroscopy (XPS). The SnAg$_2$ surface alloy was prepared by depositing Sn atoms onto the clean Ag(111) surface at elevated sample temperatures (T=550 K) using a homebuilt thermal Sn evaporator. After deposition, the sample was kept at elevated temperatures for at least $30\,$min to desorb residual non-alloyed Sn atoms. The sample coverage was confirmed after deposition by XPS via comparison of the areas of Sn3d peaks to reference data. The organic molecules were subsequently deposited from a commercial Kentax evaporator with an evaporation temperature of T$=570\,$K. The molecular film of PTCDA was deposited with a rate of approximately 10 minutes per monolayer. The sample coverage was confirmed by the C1s corelevel intensity. 

\subsection{Experimental methods}
The momentum-resolved photoemission data were obtained by momentum microscopy at the University of Kaiserslautern\cite{momentumicroscope_general,BS_IOP2019}.  
The core level and normal incidence standing waves experiments were conducted at the hard x-ray Photoemission (HAXPES) and x-ray standing wave (XSW) end station at beamline I09 at the synchrotron light source Diamond in Didcot, UK\cite{Lee.2018}. This end station is equipped with a hemispherical electron analyzer (Scienta EW4000 HAXPES), which is mounted perpendicular to the direction of the incoming photon beam. This analyzer has an angular acceptance of $\pm$ 30$^{\circ}$ and an energy resolution of $150\,$meV using the analyzer settings of our experiments ($E_{pass}$=$100\,$eV, $d_{slit}=0.5\,$mm).
The NIXSW method is a powerful tool to investigate the adsorption height of chemically inequivalent atoms above a single crystalline surface with a precision of better than $\Delta z \leq 0.04\,$\AA. A detailed description of the method can be found elsewhere \cite{WOODRUFF19981,Woodruff_2005,ZEGENHAGEN1993202}. Here, we will just give a short summary of the basic principle:
\\
For photon energies fulfilling the Bragg condition $H=k_H-k_0\,$ for a specific reflection $H=(hkl)$, a standing wave field is formed above the single crystal due to the interference between the incoming and reflecting x-ray wave field. Scanning the photon energy through the Bragg condition, one observes a phase shift of the relative complex amplitudes between the incoming and the reflected wave by $\pi$. This phase shift results in a displacement of the standing wave field by half a lattice spacing $d_{(h k l)}$ in the direction perpendicular to the Bragg planes. This shift of the standing wave field changes the photon density at any specific position $z$ above the single crystal surface as a function of the photon energy. If an atom is at a position $z$ above the single crystal surface, its XPS yield changes depending on the position of the standing wave field, e.g. the used photon energy. The experimentally observed yield curve $I(E)\,$ can be described by

\begin{equation}
\begin{split}
I(E)=1+S_R R(E)+2\vert S_I \vert \sqrt{R(E)} \\\
+F^H \cos{(\nu(E)-2\pi P^H+\Psi)},
\end{split}
\label{Eq:xsw}
\end{equation}
where $R(E)$ is the x-ray reflectivity of the Bragg reflection with its complex amplitude $\sqrt{R(E)}\,$ and phase $\nu (E)$. $S_R$, $\vert S_I \vert\,$and $\Psi\,$ are correction parameters for non-dipolar contributions to the photoemission yield. For more details, see\cite{WOODRUFF2005187,SCHREIBER2001L519,VARTANYANTS2005196,VANSTRAATEN2018106}.
 The fitting parameters of the NIXSW analysis are the coherent fraction $F^H$ and coherent position $P^H$. The coherent position $P^H$ can be interpreted as an average height of the chemical species with respect to the Bragg planes: $z=d(h\, k\, l)\times (P^H+n)$, n= 0, 1, 2,\dots , where n is the number of Bragg planes located between the surface plane and the average position of the species. The coherent fraction $F^H$ is a parameter quantifying the vertical order of this atomic species. 
 
In ideal cases, perfect vertical order corresponds to a coherent fraction $F^H=1$, while a homogeneous distribution of adsorbates between two Bragg planes would lead to a coherent fraction $F^H=0$. For single layers of molecular adsorbates on surfaces, coherent fractions of carbon, nitrogen or oxygen atoms are typically in the range of $F^H=0.8$ even for highly ordered molecular layers\cite{Kroger_2010}. This reduction of $F^H$ can be attributed to molecular vibrations and small structural defects in the molecular film which can have a significant influence on the vertical order of carbon, nitrogen and oxygen atoms due to the typically large number of these atomic species in the molecular complexes.
  
\section{Experimental Results}

\subsection{Lateral Structure}

\begin{figure}
	\centering
	\includegraphics[width=85mm]{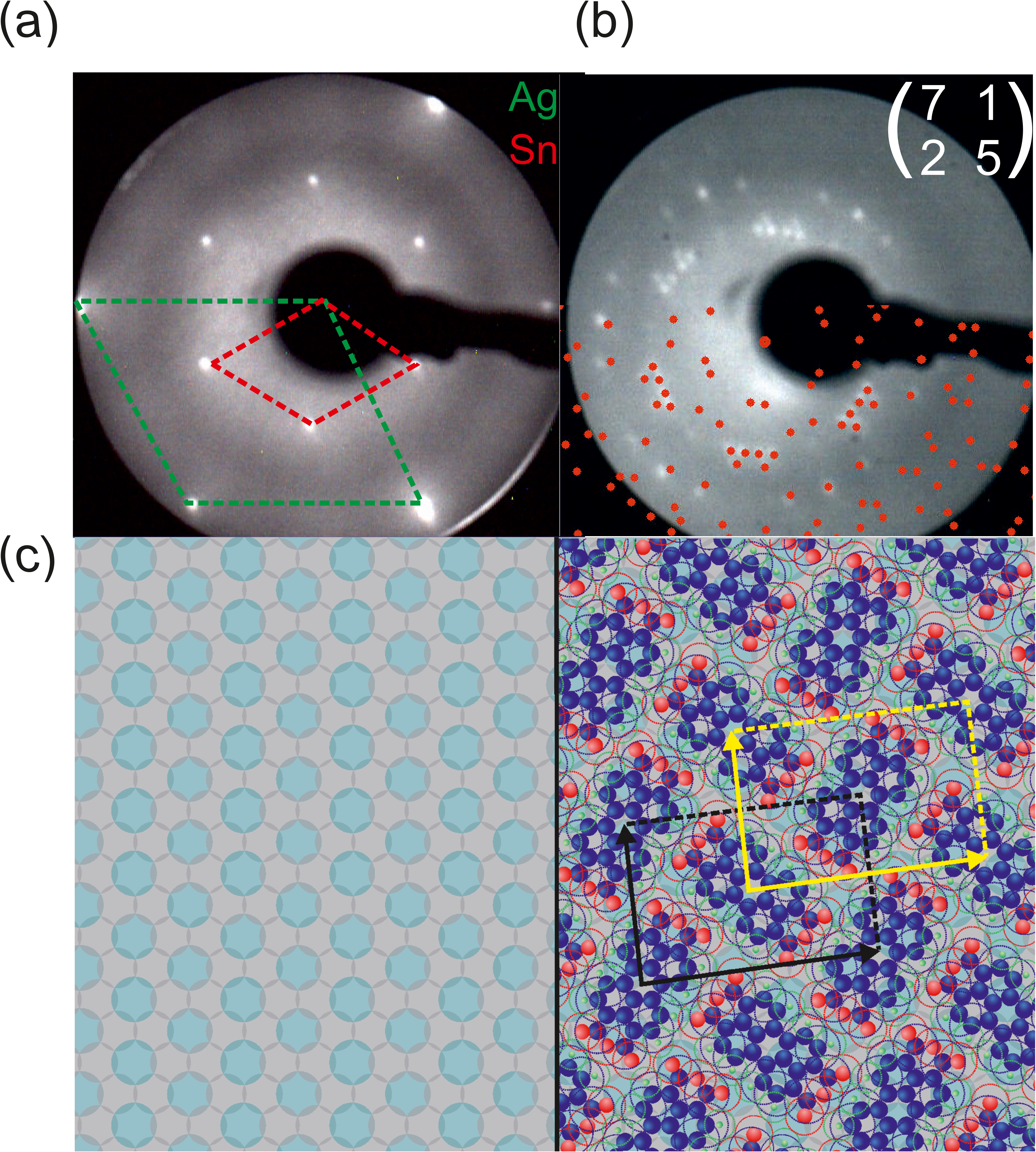} 

	\caption{
	(a) LEED image recorded for SnAg$_2$ surface alloy formed on Ag(111) (E$_{kin}=76\,$eV). The unit cell of the substrate in momentum space is sketched as green hexagon, the one of the $\left( \sqrt{3} \times \sqrt{3}\right) R30^{\circ}$ superstructure as red hexagon. (b) LEED image for a monolayer of PTCDA on SnAg$_2$ surface alloy(E$_{kin}=14\,$eV). The bottom half of the LEED image is superimposed with simulated diffraction spots of the superstructure matrix of PTCDA/Ag(111): \usebox{\smlmat}. (c) left half: bare SnAg$_2$ surface alloy, right half: structural model of the PTCDA monolayer film on the SnAg$_2$. }
\label{Fig:Fig1}
\end{figure}

We now turn to the discussion of the experimental results and start with the lateral order of the PTCDA/SnAg$_2$ interface. The bare SnAg$_2$ surface alloy has already been studied recently by Osiecki et al.\cite{OsieckiPRB2013}. Briefly, in unison with other surface alloys on fcc(111) noble metal surfaces\cite{pacile2006,AstPRL2007,Moreschini_2009,GierzPRB2010}, the SnAg$_2$ alloy exhibits a homogeneous $\left(\sqrt{3} \times \sqrt{3}\right)R30^{\circ}$ superstructure that is formed by replacing each third Ag surface atom by Sn. The corresponding low energy electron diffraction (LEED) pattern of this superstructure is shown in Fig.~\ref{Fig:Fig1}(a). The red rhombus marks the unit cell in reciprocal space of the surface alloy superstructure, the green rhombus the one of the Ag(111) bulk crystal surface. The real space structure is shown in Fig.\ref{Fig:Fig1}(c) in the left half: every Sn atom is surrounded by 6 Ag atoms.\\
After the adsorption of one monolayer of PTCDA onto the clean SnAg$_2$ surface alloy, we observe a set of new, well-defined spots in LEED (Fig.~\ref{Fig:Fig1}(b)) indicating the formation of a long-range ordered molecular overlayer. Interestingly, the LEED pattern of the PTCDA/SnAg$_2$ interface exhibits the typical arrangement of diffraction spots in double triangular structures known for PTCDA monolayer films on bare fcc(111) noble metal surfaces. 
For a quantitative analysis of the LEED data, we simulated diffraction patterns for various structures and found the best agreement between the LEED data and the simulations for the superstructure with the matrix $\left(\begin{smallmatrix}7&1\\2&5\\\end{smallmatrix}\right)$. This matrix is identical to one for the PTCDA monolayer structure on Ag(111)\cite{GLOCKLER19981,KILIAN2004359} and indicates that the overlayer is commensurate with the silver surface lattice. Consequently it is also commensurate with the $\left(\sqrt{3} \times \sqrt{3}\right)R30^{\circ}$ lattice of SnAg$_2$ alloy in higher order:
% From the superstructure matrix, we can deduce that the PTCDA overlayer exhibits a commensurate registry with respect to the silver surface lattice as well as a higher order commensurate registry with the SnAg$_2$ surface alloy. Consequently, the superstructure matrix describing a unit cell with three times larger unit cell vectors as the PTCDA monolayer structure on the SnAg$_2$ has only integer numbers:
\begin{equation*}
\label{Matrix}
    3\cdot \begin{pmatrix} \vec{A} \\ \vec{B} \end{pmatrix}_{\text{PTCDA}}
  = \begin{pmatrix} 8 & -5 \\ 7 & 8 \end{pmatrix} \cdot \begin{pmatrix} \vec{A} \\ \vec{B} \end{pmatrix}_{\text{SnAg}_2}
  = \begin{pmatrix} 21 & 3 \\ 6 & 15 \end{pmatrix} \cdot \begin{pmatrix} \vec{A} \\ \vec{B} \end{pmatrix}_{\text{Ag}}
\end{equation*}

The molecular structure of the PTCDA monolayer film on the SnAg$_2$ surface alloy is illustrated by our structural model in Fig.~\ref{Fig:Fig1}(c). This model is adopted from the one for the monolayer structure of PTCDA adsorbed on Ag(111) and exhibits two PTCDA molecules per unit cell arranged in a herringbone pattern. In discrepancy with our previous studies of adsorption of PTCDA on the PbAg$_2$ surface alloy \cite{BSPRB2016,BSPRL2016}, we do not observe a change in the lateral ordering of the molecules compared to adsorption on Ag(111). We can conclude that the Sn superstructure does not dominate the ordering process for adsorption of PTCDA molecules on the SnAg$_2$ surface alloy.

\subsection{Core Level Spectroscopy}
\begin{figure}[hb]
	\centering
	\includegraphics[width=85mm]{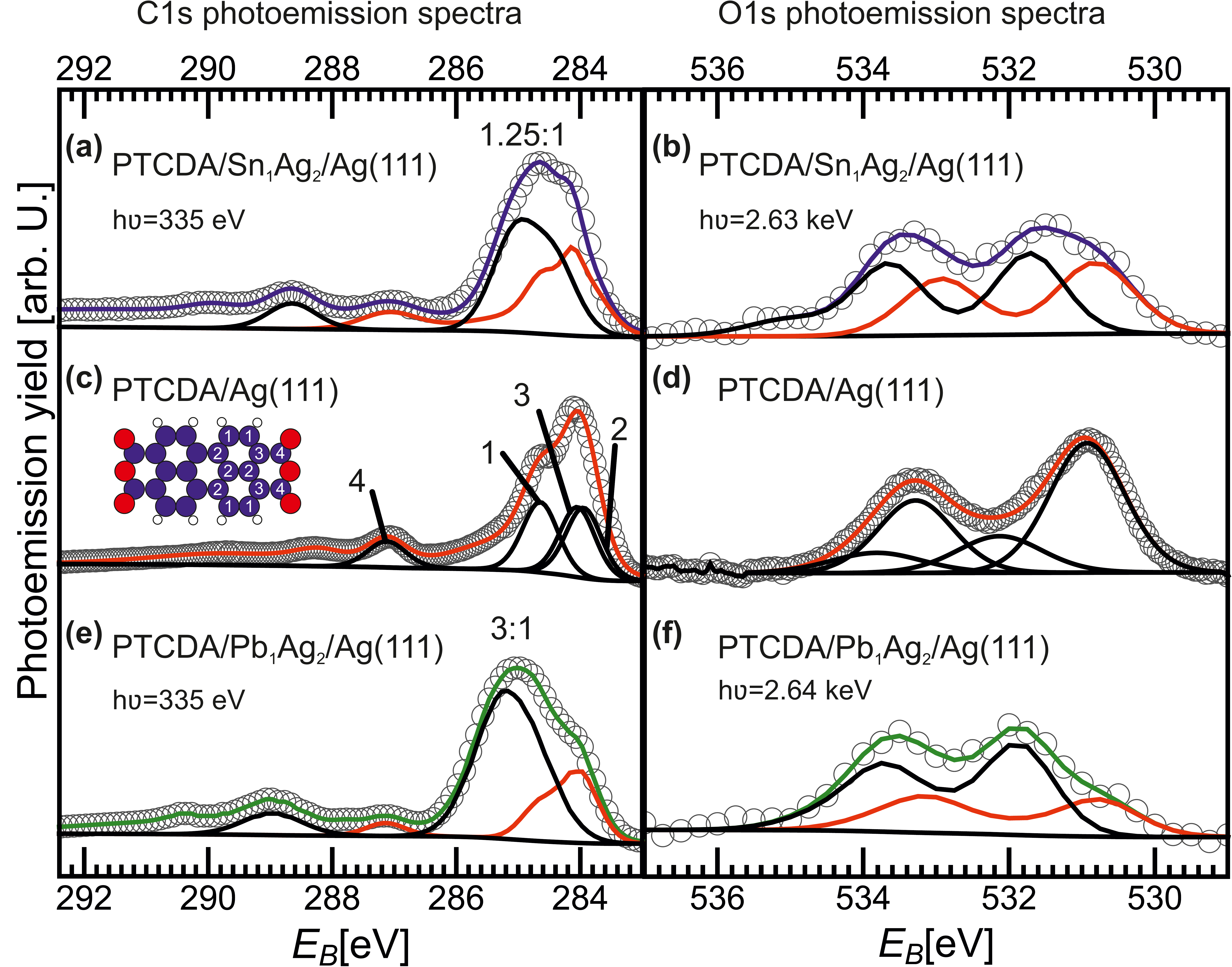} 
		\caption{The C
		1s (left) and O1s (right) core level spectra for the monolayer structure of PTCDA on (a,b) the SnAg$_2$ surface alloy, (c,d) the clean Ag(111) surface and (e,f) the PbAg$_2$ surface alloy.} 
\label{Fig:Fig2}
\end{figure}

To characterize the electronic structure of the interface, we performed core level spectroscopy studies using both the soft and hard x-ray branches of the I09 beamline. In Fig.~\ref{Fig:Fig2}(a) and (b) we show the C1s and O1s core level spectra for a layer of PTCDA adsorbed on SnAg$_2$. For comparison, we also included the corresponding C1s and O1s core level spectra for PTCDA adsorbed on the bare Ag(111) surface (Fig.~\ref{Fig:Fig2}(c),(d)) and on the PbAg$_2$ surface alloy (Fig.~\ref{Fig:Fig2}(e),(f)). 

The C1s core level spectrum of PTCDA on the SnAg$_2$ surface alloy consists of an intense main line with asymmetric lineshape and at least two side peaks with significantly lower intensity. The asymmetric shape of the main line is caused by a shoulder on its low binding energy side. This lineshape and the corresponding full width at half maximum (FWHM) clearly differs from the spectrum of PTCDA/Ag(111) (Fig.~\ref{Fig:Fig2}(c)), but reflects very well the C1s spectrum of PTCDA on the PbAg$_2$ surface alloy (Fig.~\ref{Fig:Fig2}(e)). Therefore, we adopt the fitting model developed for PTCDA/PbAg$_2$\cite{BSPRB2016,BSPRL2016} to disentangle the contributions of the chemically inequivalent atomic species to the C1s spectrum  of PTCDA/SnAg$_2$. The total C1s spectra consists of two contributions from chemically inequivalent PTCDA molecules. The fitting models are shown as red and black curve below the experimental data. The core level signature of one type of PTCDA molecule (PTCDA$_1$, red curve in Fig~\ref{Fig:Fig2}(a)) is almost identical to the one for PTCDA on the bare Ag(111) surface. It consists of four Gaussian curves which reflect the contributions of the four chemically inequivalent carbon atoms of the molecule: the C-H (1), C-C (2), C-C-O (3) and C-O (4) carbon species, indicated in the PTCDA model in  Fig.~\ref{Fig:Fig2}(c). The envelope of the best fitting result is shown as a blue solid lines in Fig.~\ref{Fig:Fig2}(a), the contributions of both types of PTCDA molecules as red (PTCDA$_1$) and black (PTCDA$_2$) solid lines. The excellent quality of the fit directly confirms the existence of two chemically inequivalent PTCDA molecules on the SnAg$_2$ surface alloy. The C1s signature of one type of PTCDA molecule (PTCDA$_2$, black curve in Fig~\ref{Fig:Fig2}(a)) is similar to that  of a PTCDA multilayer film \cite{Schoell2004} while the other is similar to adsorption on Ag(111). Our analysis reveals an almost balanced ratio between both types of PTCDA molecules of approximately 1.25:1 instead of 3:1 as observed for PTCDA/PbAg$_2$\cite{BSPRB2016}. This almost 1:1 stoichiometry is consistent with our structural model exhibiting two structurally inequivalent molecules per unit cell. In addition, the binding energy of the C-C components of both types of PTCDA molecules is shifted by $\Delta$E$=0.55\,$eV with respect to each other. A similar effect, but clearly larger shift of $\Delta$E$=0.83\,$eV was also observed within PTCDA on the PbAg$_2$ surface alloy. The core level shift between the C1s species of both types of PTCDA molecules suggests a different chemical environment for PTCDA$_1$ and PTCDA$_2$ on the SnAg$_2$ surface alloy. 

The analysis of the O1s core level data leads to a very similar conclusion. The lineshape of the O1s core level emission of PTCDA on the SnAg$_2$ surface alloy in Fig.~\ref{Fig:Fig2}(d) reveals a pronounced shoulder on its low binding energy side which is absent for PTCDA on the bare Ag(111) surface, see Fig.~\ref{Fig:Fig2}(d). In good agreement to our previous studies \cite{BSPRB2016}, such a shoulder in the O1s signature was recently observed for PTCDA on the PbAg$_2$ surface alloy\cite{BSPRL2016,BSPRB2016}, see Fig.~\ref{Fig:Fig2}(f). In analogy to the C1s core level spectrum, we analyzed the O1s lineshape with a superposition of two core level signatures of chemically inequivalent PTCDA molecules. The fit envelope is depicted as solid blue line, the corresponding contributions of both types of PTCDA molecules as red and black solid lines. For both types of molecules, we considered two chemically inequivalent oxygen species, namely the carboxylic and the anhydride oxygen species of PTCDA \cite{HauschildPRB2010}. The best fit was obtained for the model shown in Fig.~\ref{Fig:Fig2}(b).  We observe a comparable stoichiometric ratio between the signatures of both inequivalent PTCDA molecules and a relative shift of $\Delta$E$=0.9\,$eV between the carboxylic oxygen peaks of both types of molecules.

\subsection{Vertical Adsorption Configuration}
\begin{figure}
\centering
\includegraphics[width=85mm]{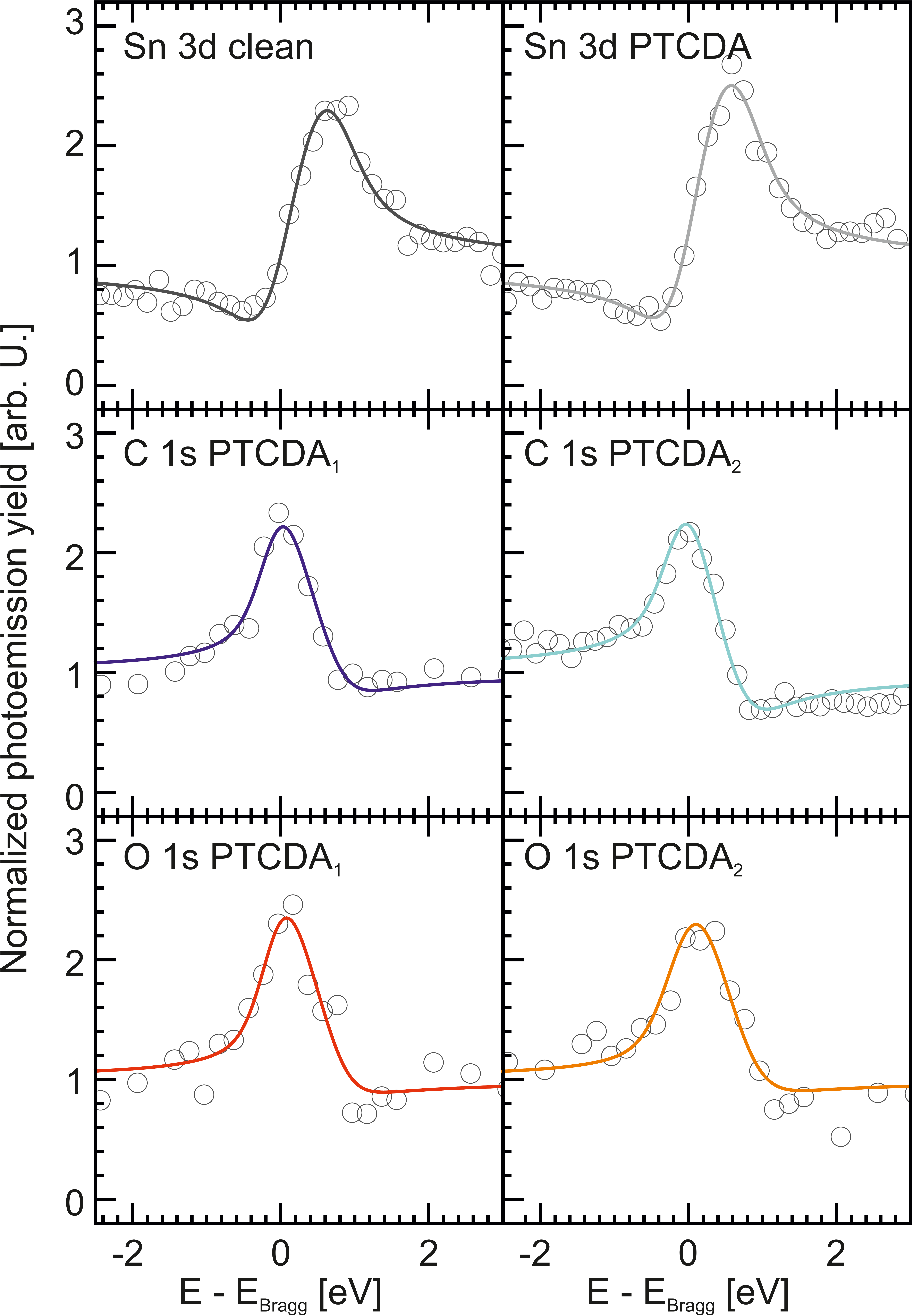}
\caption{Exemplary partial yield curves of individual NIXSW scans for all different species of the PTCDA/SnAg$_2$ interface that could be distinguished in our core level analysis.}
\label{Fig:Fig5}
\end{figure}
More insights into the origin of both types of PTCDA molecules can be obtained by investigating their vertical adsorption configurations and their vertical bonding distances at the PTCDA/SnAg$_2$ interface using the NIXSW technique.
For the NIXSW analysis, we used the core level models for C, O, and Sn discussed above. In the fitting procedure, we constrained the relative intensities of all chemically different C and O species, separately for each of the two types of PTCDA molecules. Furthermore, the binding energy positions amd FWHM of all Gaussian peaks were kept constant in the fitting. For each of the two inequivalent PTCDA molecules this results in two free fitting parameters for the adsorption heights, one for the (average height of all) carbon atoms and one for the (average height of all) oxygen atoms. While this severely reduces the structural information obtained in the NIXSW experiment, it was necessary to obtain reliable fitting results. The core level data were analyzed with the commercial software CASAXPS. Uncertainties of the photoemission yield curve was estimated by the implemented Monte Carlo error analysis\cite{MBW+13,Mercphd}. They are usually 10 \% and are omitted in the plots for better visibility. 

\begin{figure}
	\centering
	\includegraphics[width=85mm]{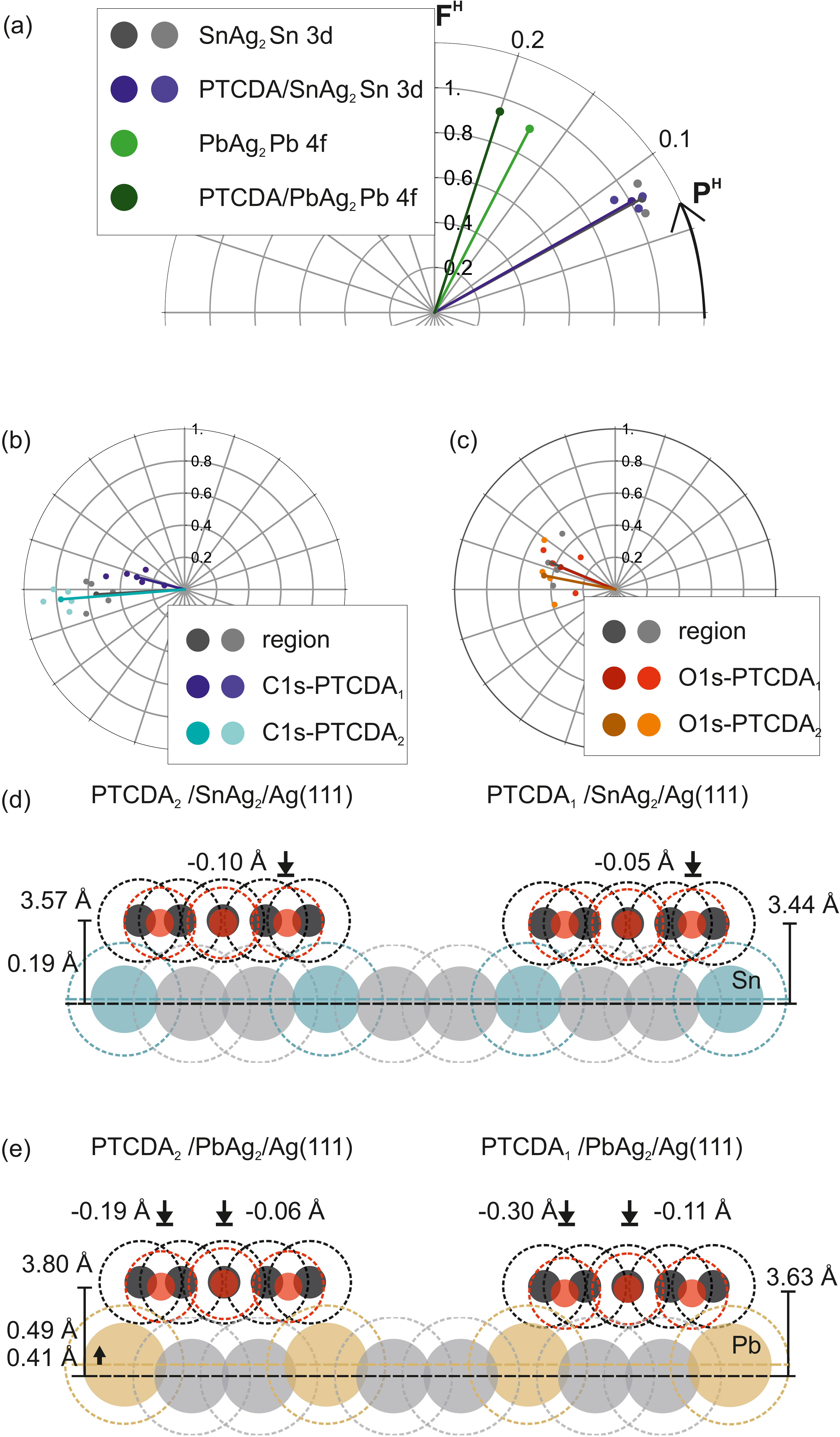} 
	\caption{(a) Argand diagram with NIXSW fitting results of the Sn 3d yield curve of the SnAg$_2$ surface alloy, prior and after the adsorption of a monolayer of PTCDA. For comparison we included the results of our previous studies of the Pb4f peak of the PbAg$_2$ surface alloy\cite{BSPRB2016} . The Argand diagram of the NIXSW fitting results of the C1s (b) and O1s (c) core levels of a PTCDA monolayer on the Sn$_1$Ag$_2$. Real scale vertical adsorption geometry model for PTCDA adsorbed on Sn$_1$Ag$_2$ surface alloy and on the Pb$_1$Ag$_2$ surface alloy are shown in (d) and (e) respectively.}
\label{Fig:Fig3}
\end{figure}

We present exemplary yield curves for every fitted atomic species in Fig.~\ref{Fig:Fig5}. The yield curves  are analyzed with the NIXSW analysis software TORICELLI \cite{toricelli,toricelli2} resulting in the fitting parameters coherent position P$^H$ and coherent fraction F$^H$ (see equation~\ref{Eq:xsw}). For each species, several scans at different sample positions were recorded and analyzed individually. The fitting results are presented in the Argand diagrams in Fig.~\ref{Fig:Fig3}: the length of the vector represents the coherent fraction, the angle included with the positive x-axis the coherent position.
The results for the Sn surface atoms are shown in the Argand diagram in Fig.~\ref{Fig:Fig3}(a), the ones for the carbon and oxygen core levels of PTCDA in Fig.~\ref{Fig:Fig3}(b) and Fig.~\ref{Fig:Fig3}(c). The fitting results of the individual scan of each chemical species are shown as colored circles. An arrow of identical color represents the average value of these fitting results for each species. The experimental uncertainty of the fitting parameters and the adsorption height of each species are calculated by the standard deviation of the individual fitting results. These averaged fitting parameters for P$^H$ and F$^H$ are summarized in Table~\ref{Tab:Tab1} together with the corresponding adsorption heights. As a reference we included the adsorption heights of all atomic species for PTCDA on the PbAg$_2$ surface alloy \cite{BSPRB2016} and on Ag(111)\cite{HauschildPRB2010}. 

\begin{table*}[t]
\begin{center}
\caption{NIXSW fitting results and corresponding adsorption heights of all chemically inequivalent species of the PTCDA/SnAg$_2$ interface. For comparison, adsorption heights for PTCDA on the PbAg$_2$ and on the Ag(111) surface are included\cite{BSPRB2016,HauschildPRB2010}. We used non-dipolar correction parameters of $\gamma$=1.06219 (0.97470) for the C1s (O1s) emission lines, respectively. The experimental geometry is given by the Bragg angle of $\theta$=86.5$^{\circ}$ (3.5$^{\circ}\,$ off normal incidence) and by $\phi$=75$^{\circ}$ (photoelectron emission angle relative to incident beam). For more details see\cite{VANSTRAATEN2018106} and \cite{toricelli} for instance. For the analysis of the Sn 3d core level, we only considered the photoemission yield in a narrow angle range close to an emission angle of $90^\circ$ to minimize the influence of non-dipolar contributions to the photoemission signal.}
\begin{tabular}{ p{0.16\textwidth} p{0.16\textwidth} p{0.16\textwidth} p{0.16\textwidth} p{0.16\textwidth} p{0.16\textwidth} } % Ausrichtung und Spaltentrennung
\hline \hline
 &$F^H$ & $P^H$ &  $D^{H}_{\mathrm{SnAg}}\,$[\AA] &  $D^{H}_{\mathrm{PbAg}}\,$[\AA] &  $D^{H}_{\mathrm{Ag}}\,$[\AA]  \\
\hline % horizontale Linie
Sn 3d$_{\mathrm{alloy}}$& $1.05 \pm 0.05$  & $0.08 \pm 0.01$ & $0.19 \pm 0.02$ & - & -  \\
Sn 3d$_{\mathrm{PTCDA}}$& $1.01 \pm 0.03$& $0.08 \pm 0.03$ & $0.19 \pm 0.008$ & - & -  \\
C1s$_{\mathrm{PTCDA,1}}$& $0.27 \pm 0.08$& $0.46 \pm 0.01$ & $3.44 \pm 0.03$  & $3.63 \pm 0.04$ & $2.86 \pm 0.01$  \\
C1s$_{\mathrm{PTCDA,2}}$& $0.69 \pm 0.05$& $0.51 \pm 0.01$ & $3.57 \pm 0.02$  & $3.80 \pm 0.04$ & $2.86 \pm 0.01$  \\
O1s$_{\mathrm{mean,1}}$& $0.33 \pm 0.07$ &$0.43 \pm 0.03$ & $3.38 \pm 0.07$ & $3.39\pm0.05$& $2.77\pm0.05$ \\
O1s$_{\mathrm{mean,2}}$& $0.41 \pm 0.04$&$0.46 \pm 0.03$ &$3.45 \pm 0.07$ & $3.65\pm0.04$ & $2.77\pm0.05$ \\
%O1s$_{\mathrm{carbox,1}}$& $0.33 \pm 0.07$& $0.43 \pm 0.03$ & $3.38 \pm 0.07$  & $3.33 \pm 0.02$ & $2.66 \pm 0.02$  \\
%O1s$_{\mathrm{carbox,2}}$& $0.41 \pm 0.04$& $0.46 \pm 0.03$ & $3.45 \pm 0.07$  & $3.61 \pm 0.02$ & $2.66 \pm 0.02$  \\
%O1s$_{\mathrm{anhy,1}}$& $0.33 \pm 0.07$& $0.43 \pm 0.03$ & $3.38 \pm 0.07$  & $3.52 \pm 0.09$ & $2.98 \pm 0.08$  \\
%O1s$_{\mathrm{anhy,2}}$& $0.41 \pm 0.04$& $0.46 \pm 0.03$ & $3.45 \pm 0.07$  & $3.74 \pm 0.05$ & $2.98 \pm 0.08$  \\

\hline \hline
\end{tabular}
\label{Tab:Tab1}
\end{center}
\end{table*}
For the bare SnAg$_2$ surface alloy (before adsorption of PTCDA), we find a coherent fraction of F$^H_{Sn}=1.05\pm   0.05  $ and a coherent position of P$^H_{Sn}=0.08 \pm 0.01$. The experimental uncertainties are rather small, as reflected by the marginal scattering of the fitting results of the individual NIXSW scans in the Argand diagram in Fig.~\ref{Fig:Fig3}(a). The corresponding adsorption height of Sn is $0.19\pm 0.02\,$\AA$\,$ with respect to the surface plane of the Ag atoms, i.e., the Sn atoms show a small vertical relaxation from the surface layer towards the vacuum. Interestingly, our experimental findings cannot confirm recent theoretical predictions for a small vertical inwards relaxation ($\Delta z=-0.04\,$\AA) of the Sn atoms of a SnAg$_2$ surface alloy towards the Ag bulk \cite{OsieckiPRB2013}. Moreover, the vertical relaxation of the Sn atoms from the Ag surface plane is comparable to the relaxation of Sb, but significantly smaller than the relaxation of other surface alloy atoms such as Bi or Pb\cite{GierzPRB2010}. We attribute this difference to the smaller atomic size of Sn or Sb compared to Bi and Pb. 

The adsorption of one layer of PTCDA on the SnAg$_2$ surface alloy does not change the vertical positon of the Sn atoms (within experimental uncertainties), as we find almost identical coherent positions and coherent fractions before and after adsorption of PTCDA. The missing vertical relaxation of the Sn atoms represents a significant difference to our recent study of PTCDA on PbAg$_2$ surface alloy. For the latter system, the formation of local $\sigma$-bonds between the PTCDA molecule and the surface atoms resulted in an additional vertical relaxation of the Pb surface alloy atoms of $\Delta z= 0.08 \pm 0.01\,$\AA \cite{BSPRB2016}. Such bonds are clearly absent for the PTCDA/SnAg$_2$ interface as indicated by the Sn adsorption heights obtained in this study. 

 The NIXSW fitting results for the carbon and oxygen atoms of both inequivalent PTCDA molecules PTCDA$_1$ (monolayer-like) and PTCDA$_2$ (bulk-like) are shown in the Argand diagram Fig.~\ref{Fig:Fig3}(b) and (c). The scattering of the fitting results for both elements is significantly larger compared to the one for the Sn surface alloy atoms resulting in larger experimental uncertainties of the vertical positions of both types of PTCDA molecules. In addition, the coherent fractions of the carbon and oxygen atoms of both PTCDA molecules are clearly different. The coherent fractions of F$^H_{C1s}=0.69\pm 0.05$ and F$^H_{O1s}=0.41\pm 0.04$ of PTCDA$_2$ are comparable with coherent fractions for flat lying PTCDA molecules on other noble metal surfaces. However, the corresponding fractions of PTCDA$_1$ of F$^H_{C1s}=0.27\pm 0.08$ and F$^H_{O1s}=0.33\pm 0.07$ are significantly lower than typical values for flat lying molecules. We suspect that the reason for this lower fraction is not a strong bending or tilt of the PTCDA$_1$ molecules, but originates from a small contribution of the photoemission yield of the PTCDA$_2$ molecules to the one of PTCDA$_1$ due to the energetic overlap between the individual peaks of both PTCDA species in the C1s and O1s core level data. This reduces the coherent fraction of PTCDA$_1$ due to the different adsorption heights of both types of molecules on the surface alloy. Despite this challenge in the data analysis of PTCDA$_1$, the fitting results of PTCDA$_1$ still reflect the structural properties of this type of PTCDA molecule. This conclusion is supported by the significantly different coherent positions P$^H$ of both types of PTCDA molecules obtained in the NIXSW analysis. 

For a further discussion of the NIXSW results, the vertical adsorption geometry of both PTCDA molecules is shown in a true scale model in Fig.~\ref{Fig:Fig3}(d). As a reference, a similar model of both types PTCDA molecules on the PbAg$_2$ surface alloy is shown in Fig.~\ref{Fig:Fig3}(e). The PTCDA molecules are drawn in a schematic side view along their long molecular axis. The colored circles indicate the (empiric) atomic radii while the dashed circles show the corresponding van-der-Waals-radii \cite{vdwnew}.

Upon adsorption on the SnAg$_2$ surface alloy, both types of PTCDA molecules exhibit  almost no bending of the oxygen end groups towards the interface. While the average adsorption position of the oxygen atoms of the anhydride endgroups is slightly below the carbon backbone ($\Delta z_{PTCDA,1}=-0.06\pm 0.07\,$\AA, and $\Delta z_{PTCDA,2}=-0.12\pm 0.07\,$\AA), this difference is barely significant within the experimental uncertainty. This observation is in clear contrast to the vertical adsorption geometry of PTCDA on PbAg$_2$\cite{BSPRB2016} surface alloy as well as on all low index silver surfaces\cite{HauschildPRB2010,MBW+13,Bauer_PRB2012}. On these surfaces, at least one chemical oxygen species (typically the carboxylic oxygen atoms) bends down towards the surface atoms. This vertical distortion of the PTCDA molecules is the geometric signature of the formation of local, $\sigma$-like bonds between the oxygen atoms of PTCDA and the surface. In contrast, the flat adsorption geometry of both inequivalent PTCDA molecules on the SnAg$_2$ surface alloy indicates the absence of such local bonds between the oxygen atoms and the Sn or Ag surface atoms. Such $\sigma$-bonds were identified as the microscopic origin of the vertical relaxation of the Pb atoms of the PbAg$_2$ surface alloy after the adsorption of PTCDA.

The carbon backbone of both PTCDA molecules is located at slightly different adsorption heights of D$^H_{PTCDA,1}=3.44\pm 0.03\,$\AA$\,$ and D$^H_{PTCDA,2}=3.57\pm 0.02\,$\AA. This difference is identical to the difference in adsorption heights of the two chemically inequivalent PTCDA molecules on the PbAg$_2$ surface alloy. However, the overall adsorption height of PTCDA on SnAg$_2$ is $\approx 0.2\,$\AA$\,$ smaller than on the PbAg$_2$ surface. This can be attributed to the significantly larger vertical buckling and atomic radius of the Pb surface atoms compared to the Sn atoms in their respective surface alloys on Ag(111). 
A better representation of the distances between two atoms is given by the normalized bonding distance which is calculated by
\begin{equation}
d^{N}_{A-B}=\frac{d_{A-B}}{d_{VdW}^{A}+d_{VdW}^{B}} .
\end{equation}
The normalized bonding distance is a very good indicator for the interaction strength between two atoms A and B. While $d^{N}_{A-B}\approx 1\,$ represents very weak interaction, e.g. maximal physisorption, smaller radii due to partial overlapp of the atomic van-der-Waals radii indicate stronger interaction. The results  for the interfaces formed between PTCDA with SnAg$_2$, PbAg$_2$ and Ag(111) are summarized in Table~\ref{Tab:Tab2}. We obtain higher normalized bonding distances in the SnAg$_2$ system compared to both PbAg$_2$ and Ag(111) for all atomic species.  

\begin{table}[b]
\begin{center}
\caption{Normalized bonding distances for PTCDA monolayer films on the SnAg$_2$, PbAg$_2$ and Ag(111) surfaces. Van der Waals radii were taken from \cite{vdwnew}: $r_C=1.77\,$\AA , $r_O=1.50\,$\AA , $r_{Sn}=2.42\,$\AA , $r_{Pb}=2.54\,$\AA $\,$and $r_{Ag}=2.53\,$\AA .}
\begin{tabular}{ c c c c } % Ausrichtung und Spaltentrennung
\hline \hline
 &  $d^{N}_{\mathrm{SnAg}}\,$[\%] &  $d^{N}_{\mathrm{PbAg}}\,$[\%] &  $d^{N}_{\mathrm{Ag}}\,$[\%]  \\
\hline % horizontale Linie

C1s$_{\mathrm{PTCDA,1}}$& $78 \pm 1.4$& $73 \pm 1.2$ & $68 \pm 1.5$ \\
C1s$_{\mathrm{PTCDA,2}}$& $81 \pm 1.2$& $77 \pm 0.7$ & $68 \pm 1.5$ \\
O1s$_{\mathrm{carbox,1}}$& $82 \pm 2.6$& $70 \pm 0.7$ & $68 \pm 2.3$ \\
O1s$_{\mathrm{carbox,2}}$& $84 \pm 2.6$& $77 \pm 0.7$ & $68 \pm 2.3$   \\
O1s$_{\mathrm{anhy,1}}$& $82 \pm 2.6$& $75 \pm 2.5$ & $76 \pm 4.1$   \\
O1s$_{\mathrm{anhy,2}}$& $84 \pm 2.6$& $80 \pm 1.5$ & $76 \pm 4.1$  \\

\hline \hline
\end{tabular}
\label{Tab:Tab2}
\end{center}
\end{table}
Altogether, our investigation of the vertical adsorption geometry shows that surface alloying of the Ag(111) surface with Sn atoms suppresses the chemical interaction between PTCDA and the Ag(111) surface. In particular, we do not observe any vertical relaxation of the Sn surface alloy atoms which indicates the absence of local, $\sigma$-like bonds between the molecule and the Sn alloy atoms. In the following, we will explore the consequence of the apparently physisorptive molecule-surface bonding for the surface band structure of the alloy and the interfacial energy level alignment at the interface.

\subsection{Electronic valence band structure}

\begin{figure}
	\centering
	\includegraphics[width=85mm]{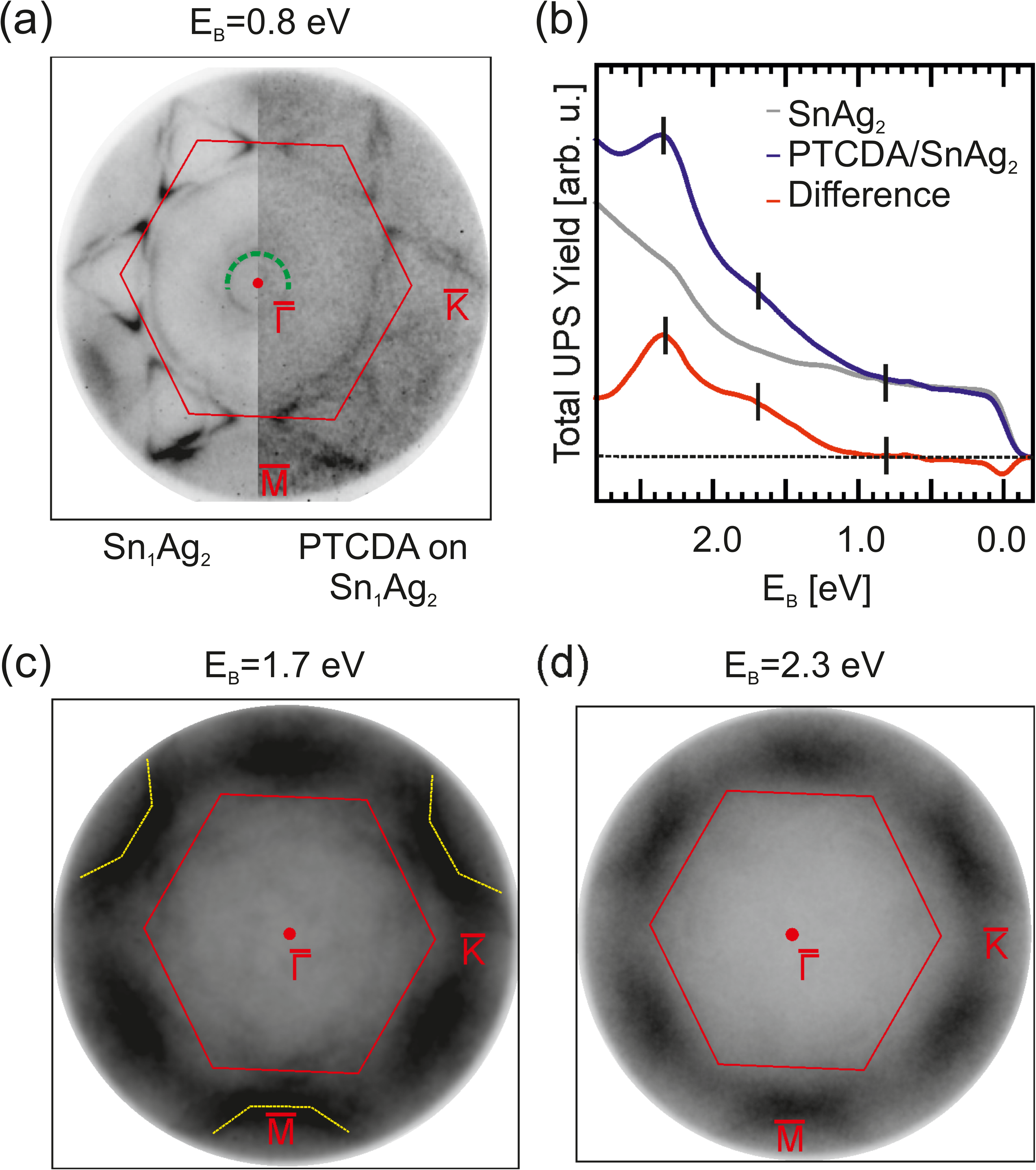} 
	\caption{(a) Constant energy map at $0.8\,$eV binding energy for the bare SnAg$_2$ surface alloy (left) and after adsorption of one layer of PTCDA (right). The hybrid interface state (indicated by a green ring) is not altered upon adsorption of PTCDA. (b) Total UPS yield of the electronic valence structure of PTCDA/SnAg$_2$ (blue), SnAg$_2$ (gray) and the difference signal (red). We observe no molecular feature at $E_F\,$, but two molecular features at E$_B=1.7\,eV\,$ and E$_B=2.3\,eV$. (c, d) Symmetrized constant energy maps extracted at E$_B=1.7\,eV$ and E$_B=2.3\,eV\,$ respectively, i.e., at the binding energy of the molecular features in the UPS spectrum. }
\label{Fig:Fig4}
\end{figure}

The modification of the SnAg$_2$ surface band structure upon adsorption of PTCDA can be determined directly  by momentum-resolved photoemission spectroscopy with vacuum-ultraviolet radiation. 
The band structure of the bare SnAg$_2$ is dominated by a hybrid surface state with a hole-like parabolic dispersion, which is located at the $\bar{\Gamma}$-point of the surface Brillouin zone\cite{OsieckiPRB2013}. In the momentum resolved photoemission data shown in Fig.~\ref{Fig:Fig4}(a), left side, this hybrid surface state appears as ring-like feature in the constant energy map (CE map) recorded at E$_B=0.8\,$eV (indicated by the green ring). Upon adsorption of PTCDA, no modification of the emission pattern or the band dispersion of this hybrid surface state is observed in the occupied valence band structure as exemplarily shown in the CE map in Fig.~\ref{Fig:Fig4}(a), right side. This CE map was obtained for the same energy as the corresponding map of the bare surface alloy on the left. 

These findings indicate that the adsorption of PTCDA does not significantly affect the surface band structure, in contrast to our recent findings for PTCDA on the PbAg$_2$ surface alloy, for which the formation of $\sigma$-like bonds between PTCDA and the surface was observed\cite{BSPRL2016}. The PTCDA/SnAg$_2$ resembles, qualitatively, the adsorption scenario observed for CuPc on the PbAg$_2$ surface alloy, a molecular adsorbate system without local $\sigma$-like bonds at the interface.

To characterize the interfacial energy level alignment and to reveal a possible adsorption-induced charge redistribution across the interface, we have recorded the total photoemission yield in the valence band structure, as shown in Fig.~\ref{Fig:Fig4}(b) as solid blue line. As a reference, the total photoemission yield of the bare SnAg$_2$ surface alloy is shown as the gray solid line, and the difference signal in red. 

Most importantly, the photoemission spectrum does not reveal any molecular feature in the vicinity of the Fermi edge, which would indicate an occupation of a LUMO derived state. Such a state is usually observed for PTCDA on Ag(111) and was also found on the PbAg$_2$ surface alloy. This state becomes (at least) partly occupied due to an interfacial charge transfer from the surface into the molecule. It can be thus regarded as the spectroscopic signature of a partial chemical interaction between the molecule and the surface mediated by a delocalized $\pi$-bond at the interface\cite{Romaner_2009}. The absence of such a state further supports our previous conclusion of a mere physisorption of PTCDA on the SnAg$_2$ surface alloy.

The two photoemission features at E$_{B}=1.7\,$eV and E$_{B}=2.3\,$eV can be assigned to the highest occupied molecular orbital (HOMO) of the two inequivalent PTCDA molecules, in analogy to the finding for PTCDA on the PbAg$_2$ surface alloy. 

This assignment is further confirmed by the CE maps of both HOMO features in Fig.~\ref{Fig:Fig4}(c) and (d). Both CE maps reveal the characteristic emission pattern of the PTCDA HOMO on a threefold symmetric fcc(111) surface with six well-defined maxima located on a ring with $r\approx 1.7\,$\AA$\,$ in momentum space\cite{BSEPL2012}. The feature at E$_{B}=1.7\,$eV corresponding to the HOMO of PTCDA$_1$ is slightly less well defined due to an overall lower intensity. For this reason, the sp-bands of the surface alloy still contribute to the intensity distribution in this constant energy map. These sp-bands are highlighted by a dashed yellow line and lead to a distortion of the expected elliptical shapes of the molecular emission pattern due to the superposition of the molecular orbital feature and substrate bands. Besides this distortion, the momentum distribution is similar to the momentum distribution of the CE map at E$_{B}=2.3\,$eV.

\section{Discussion}

Our comprehensive investigation of the structural and electronic properties of PTCDA on the SnAg$_2$ surface alloy has revealed clear signatures of a non-chemical interaction between PTCDA and the underlying surface alloy. We find a flat adsorption geometry of PTCDA as well as the absence of any charge transfer across the interface. In addition, the vertical buckling  of the Sn surface alloy atoms and the surface band structure of the SnAg$_2$ surface alloy are fully preserved after the adsorption of PTCDA. This is significantly different for PTCDA adsorbed on the formally similar PbAg$_2$ surface alloy where the formation of local $\sigma$-bonds between PTCDA and the Pb atoms results in a vertical relaxation of the Pb atoms coinciding with a strong modification of the surface alloy band structure. 
From this comparison, it can be concluded that adsorption-induced modification of the structural and electronic properties of the surface alloy crucially depends on the presence or absence of $\sigma$-bonds between the molecular adsorbates and the surface alloy. The most obvious difference between both adsorbate systems is the significantly larger intrinsic vertical buckling of the clean alloy atoms for the PbAg$_2$ ($\Delta z_{Pb}^{clean}=0.42\pm 0.02\,$\AA ) compared to the SnAg$_2$ surface alloy ($\Delta z_{Sn}^{clean}=0.19\pm 0.01\,$\AA ). These different vertical positions of the alloy atoms result in different normalized bonding distances  with the oxygen atoms of PTCDA. These, on turn, are eventually responsible for the formation of local $\sigma$-bonds at the PTCDA-surface alloy interface. While the vertical position of the alloy atoms can explain the different bonding of PTCDA to Sn- and Pb-based surface alloys on the Ag(111) surface, it cannot explain the recent findings of Cottin et al. \cite{Cottin_2014} for PTCDA on the BiAg$_2$ surface alloy. They observed no modification of the BiAg$_2$ surface band structure although the vertical buckling of the Bi atoms of $\Delta z_{Bi}=0.65\pm 0.10\,$\AA \cite{GierzPRB2010} even exceeds the one of Pb. The difference between Bi- and Pb-based alloys can be attributed to the different occupation of their hybrid surface state with $p_z$ character. While the $p_z\,$hybrid surface states of the bare SnAg$_2$\cite{OsieckiPRB2013} and the PbAg$_2$\cite{pacile2006} surface alloy are only partially occupied, the Bi-Ag $p_z\,$ hybrid surface state is completely filled\cite{AstPRL2007} due to one additional valence electron of Bi compared to Sn and Pb\cite{Mirhosseini_2010}. This different population prevents the formation of $\sigma$-bonds between PTCDA and the BiAg$_2$ surface alloy atoms, which requires an effective charge transfer between the oxygen groups of PTCDA and the unoccupied (surface) states of the surface alloy\cite{Romaner_2009}. In contrast, such a charge transfer is - in general - possible for surface alloys with partially occupied surface states such as the SnAg$_2$ and the PbAg$_2$ surface alloys, and can hence results in the formation of local $\sigma$-bonds for these surfaces. 

Therefore, we can propose three empiric rules that define the necessary requirements of  the formation of local $\sigma$-bonds between molecular complexes and surface alloy atoms:  the existence of (i) functional and reactive molecular groups, (ii) only partially occupied hybrid surface states with p$_z$ character, and (iii) a sufficient intrinsic vertical relaxation of the surface alloy atoms. These empiric rules represent the foundation for band structure engineering of surface alloys by the formation of tailored molecule-alloy surface bonds.

\section{Summary}

In this work, we have investigated the structural and electronic properties of the interface formed between PTCDA and the SnAg$_2$ surface alloy. The SnAg$_2$ surface alloy exhibits the well-known $\left(\sqrt{3} \times \sqrt{3} \right)R30^{\circ}$ superstructure and a small but significant vertical relaxation of the Sn surface alloy atoms of $0.19\pm 0.01\,$\AA$\,$ with respect to the plane of the Ag surface atoms. The adsorption of PTCDA on the Sn-Ag surface alloy does not affect the structural (vertical position of the Sn atoms) and electronic properties (surface band structure) of the SnAg$_2$ surface alloy. This finding is in contrast to the behavior of PTCDA on a PbAg$_2$ surface alloy for which the molecular adsorption modifies both the surface band structure and the vertical relaxation of the Pb alloy atoms. This apparent contradiction could be resolved by the investigation of the vertical bonding distances and the charge redistribution at the interfaces. PTCDA adsorbs in a completely flat adsorption geometry on the SnAg$_2$ surface alloy without any structural indications for the formation of local $\sigma$-like bonds to the surface. In addition, no signal of interfacial charge transfer was observed. All these findings point to the mere physisorption of PTCDA/SnAg$_2$, and not to a chemical interaction with $\sigma$-like bonds as was observed for the bare and the Pb alloyed Ag(111) surfaces. In conjunction with previous findings for PTCDA on different surface alloys\cite{BSPRB2016,BSPRL2016,Cottin_2014}, we propose three empiric rules that define for the modification of surface alloys by $\sigma$-like bonding with molecular adsorbates:  (i) functional and reactive molecular groups, (ii) only partially occupied hybrid surface states with p$_z$ character, and (iii) a sufficient vertical relaxation of the surface alloy atoms. In this way, our work has uncovered the necessary ingredients for engineering the surface band structure of binary metallic surfaces and adsorbate systems by the formation of tailored molecule-surface bonds.

\section{Acknowledgements}
The research leading to these results was funded by the Deutsche Forschungsgemeinschaft (DFG, German Research Foundation) - TRR 173 - 268565370). M.F. and C.K also acknowledge financial support from the DFG via SFB 1083, B.S. from the Graduate School of Excellence Mainz (Excellence initiative DFG/GSC 266). L.L.K. thankfully acknowledges financial support from the Carl Zeiss Stiftung. M.C. acknowledges funding from the European Research Council (ERC) under the European Union's Horizon 2020 research and innovation programme (Grant Agreement No. 725767-hyControl). Moreover, we thank Diamond Light Source for access to beamline I09 (through proposal SI13773-1). We are very grateful for the support by the beamline staff during the experiment, in particular by P. K. Thakur, D. A. Duncan, T.-L. Lee, and D. McCue.

%\bibliography{Literature}

\end{document}